# Photonic MOS Based on "Optical Property Inversion"

Zhaolin Lu[*,†] and Kaifeng Shi[*]

[*]*Microsystems Engineering PhD Program*
[†]*Department of Electrical and Microelectronic Engineering*
*Rochester Institute of Technology, Rochester, New York, 14623, USA*
zhaolin.lu@rit.edu

Most dielectric materials have very weak electro-optic properties, whereas the optical properties of some plasmonic materials may be greatly tuned, especially around their plasma frequency, where dielectric constant is transiting between positive ("dielectric state") and negative ("metallic state") values. In this report, we will review some of our recent work on electro-optical modulation and introduce a new concept, photonic MOS based on "optical property inversion". This concept may provide inspiration for the development of nanophotonic devices. Some of the material is adapted from unfunded NSF proposals of the first author. Throughout this report, "electric dielectric constant", $\varepsilon$, refers to material dielectric constant in the DC or radio frequency (RF) regime; "optical dielectric constant", $\epsilon$, represents material dielectric constant in the near-infrared regime.

## 1. WEAK EO PROPERTIES OF CONVENTIONAL DIELECTRICS

Electro-optical (EO) modulators up-convert electronic signals into high bit-rate photonic data. In addition to direct modulation of lasers, EO modulators can be classified into (i) phase modulation based on EO effect or free-carrier injection [1, 2], or (ii) absorption modulation based on Franz-Keldysh effect or quantum-confined Stark effect [3- 7]. Due to the poor EO properties of regular materials [8, 9], a conventional EO modulator has a very large footprint [10]. Recent years have witnessed breakthroughs in the development of micro EO modulators [11- 20 ]. On-chip optical interconnects require ultrafast EO modulators at the nanoscale with low insertion loss. This technical bottleneck has existed for decades and numerous devices have been reported. The performance of most devices has reached their physical limits restricted by materials based on some sophisticated techniques. We anticipate that the bottleneck may not be well resolved based on well-known materials; instead, an innovative approach should be followed. Most of previous effort was focused on the exploration of the EO properties of dielectrics owing to their low optical absorption for waveguide applications, whereas the EO properties of metals and semimetals, or generally "plasmonic materials", have been relatively overlooked.

Generally speaking, dielectrics (including insulators and semiconductors) have positive dielectric constants, whereas metals have negative dielectric constants. Research on metamaterials has shown that the dielectric constants of materials can be engineered to be almost any arbitrary value (positive, zero, or negative). One example is epsilon-near-zero (ENZ) materials [21-23]. Optically, ENZ is a critical point, where the optical property is transiting between "dielectric state" (with positive dielectric constant) and "metallic state" (with positive dielectric constant). Any slight change in the dielectric constant may result in "optical property inversion" or the switch between these two distinct states, which is very promising, according to recent works on the optical and EO properties of transparent conducting oxides (TCOs) to be detailed in next section.

## 2. DRUDE MODEL FOR CONDUCTIVE MEDIA

The optical dielectric constant $\epsilon$ of a TCO is determined by its free carrier concentration, which is higher than that of semiconductors but lower than that of metals. The effect of free carriers on an optical material can be approximated by the Drude model,

$$\epsilon = \epsilon' + j\epsilon'' = \epsilon_\infty - \frac{\omega_p^2}{\omega(\omega + j\gamma)} \tag{1}$$

where $\epsilon_\infty$ is the high frequency dielectric constant, $\omega_p$ is the plasma frequency, and $\gamma$ is the electron damping factor. $\epsilon'$ and $\epsilon''$ are real and imaginary parts of optical dielectric constant, respectively. Roughly speaking, when $\omega > \frac{\omega_p}{\sqrt{\epsilon_\infty}}$, $\epsilon' > 0$, the material functions as a dielectric; when $\omega < \frac{\omega_p}{\sqrt{\epsilon_\infty}}$, $\epsilon' < 0$, the material functions as a metal. The plasma frequencies of most metals are located in the ultraviolet or deep ultraviolet regime owing to their high carrier concentration. Note plasma frequency $\omega_p = \sqrt{\frac{Nq^2}{\epsilon_o m^*}}$, depending on carrier concentration $N$, and the effective electron mass $m^*$. $q$ represents the elementary charge. To make $\omega_p$ located in the near infrared (NIR) regime, the carrier concentration should reduce to $10^{20} \sim 10^{22}$ cm$^{-3}$, which coincides that of TCOs.

Considerable effort has been focused on TCOs as the plasmonic material for NIR applications [24-28]. Noginov, *et al*. [29] and Naik, *et al*. [30,31] did comparative studies independently, and showed that aluminum-doped zinc oxide (AZO) has even smaller absorption than indium tin oxide (ITO) in the NIR regime. More importantly, several recent experiments showed that the dielectric constant of TCOs in the accumulation layer of a metal-oxide-TCO (MOT) structure can be tuned in a large range by electrical gating [32-35]. Basically, the MOT structure functions as a parallel plate capacitor. When a bias is applied across the oxide layer, surface charge can be excited and the induced charge can alter the optical properties at the oxide-TCO interface. According to the bias polarity and strength, a MOS structure may work in three well-known modes, namely accumulation, depletion, and inversion. See Fig. 1. As the key element for very-large-scale integration (VLSI), the operation of a MOSFET is based on "carrier inversion", where electrical gating-induced surface electrons (or holes) function as the dominant carriers in a *p*-type (or *n*-type) of semiconductor channel.

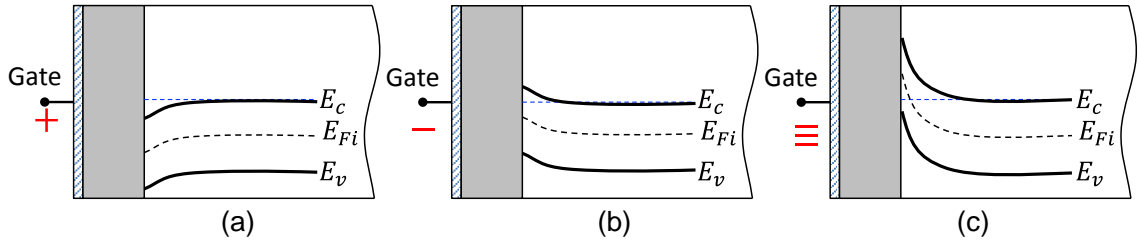

Figure 1. The three electronic operation models of a MOS structure: (a) accumulation; (b) depletion; (c) inversion.

## 3. PHOTONIC MOS BASED ON "OPTICAL PROPERTY INVERSION"

Similarly, the electrical gating-induced surface charge may also result in "optical property inversion" in a MOT structure, where the dielectric constant of TCO in the charged layer is switched between positive ("dielectric state") and negative ("metallic state") values. This is possible because the electrical gating-induced charge, *ΔN*, shifts the plasma frequency of the involved TCO, $\omega_p = \sqrt{\dfrac{Nq^2}{\epsilon_o m^*}}$, in the charged layer. For a given working frequency $\omega$, the plasma frequency $\omega_p$ shift may change the sign of the optical dielectric constant $\epsilon = \epsilon_\infty - \frac{\omega_p^2}{\omega(\omega+j\gamma)}$. In a simple formula,

$$\Delta N \rightarrow \Delta\omega_p \rightarrow \Delta\epsilon: \begin{cases} \text{Dielectric (transparent)} & \text{if } \epsilon' > 0 \\ \text{Metallic (opaque)} & \text{if } \epsilon' < 0 \end{cases} \quad (2)$$

The change in optical property may be significant because dielectric tends to be transparent, and metal, opaque. Herein, we propose "photonic MOS", i.e. MOT, based on "optical property inversion".

## 4. THEORY ON "OPTICAL PROPERTY INVERSION"

Consider an MOT structure made of three layers, namely aluminum, 10-nm HfO$_2$, and ITO, respectively. Let us first ignore the interface traps at oxide-TCO boundary for now, which may probably play an important role on the charge distribution. Also, assume the surface potential $\phi_s = 0$, i.e. "flat band", when no bias is applied. The electrical permittivity values of HfO$_2$ and ITO are assumed to be $\varepsilon_{ox} = 25\epsilon_o$ and $\varepsilon_t = 9.3\epsilon_o$. $\epsilon_o$ is the permittivity of free space. Also assume the carrier concentration of the ITO to be $N_d = 4 \times 10^{20}$ cm$^{-3}$.

### 4.1. Mode 1: Depletion (Dielectric State)

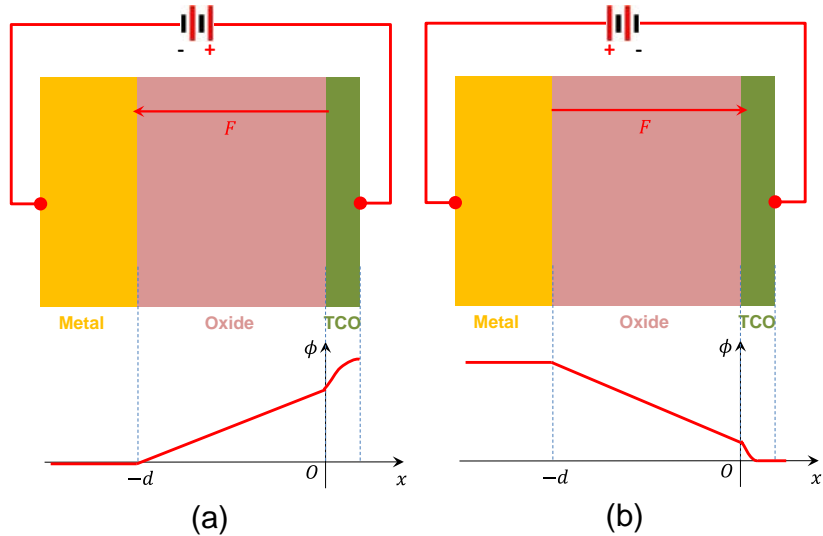

**Figure 2.** (a) The illustration of the MOT structure working at the depletion mode. (b) The illustration of the MOT structure working at the accumulation mode.

When a negative bias is applied, the MOT structure works at the depletion mode. See Fig. 2(a). If a negative bias $-V_b$ is applied across the HfO$_2$ layer (the metal layer is negatively biased), the excited surface charge on the metal side will be $Q = -\frac{\varepsilon_{ox}A}{d}V_b = A\varepsilon_{ox}F$, or

$$-qN_s = \varepsilon_{ox}F \tag{3}$$

where $d$ is the thickness of the oxide layer, $F$ is the applied electric field across the oxide layer, and $-qN_s$ is the surface charge density at the metal-oxide boundary. If $V_b = -20$ V, the excited surface charge will be $-qN_s = -\varepsilon_{ox}F = -0.443$ C/m$^2$ or $N_s = 2.76 \times 10^{18}$ m$^{-2}$ according to Eq. (3). Meanwhile, a depletion layer is formed in the TCO layer. The depletion width can be calculated as

$$w_d = \frac{N_s}{N_d} = 6.91 \text{ nm}$$

Yet, the overall voltage across the MOT structure is actually larger than that across the oxide layer:

$$V_b = \frac{1}{2}\left(\frac{qN_d}{\varepsilon_t}\right)w_d^2 + \frac{qN_d}{\varepsilon_{ox}}w_d d = 38.6 \text{ V}$$

If the TCO thickness is less than $w_d$, the whole TCO layer will be depleted. In addition, holes may be excited in TCO for low speed operation.

### 4.2. Mode 2: Accumulation (Metallic State)

An accumulation layer is formed when a positive bias is applied as shown in Fig. 2(b). Let us revisit the operation of MOS devices [36]. The distribution of the charge as a function of depth, $x$, can be found using Poisson's equation

$$\frac{d^2\phi(x)}{dx^2} = -\frac{\rho(x)}{\varepsilon_t} = -\frac{q}{\varepsilon_t}(p - n + N_d - N_a)$$

where $\phi(x)$ is the potential in the TCO as a function of depth. In the electron accumulation layer formed in $n$-type TCO, $n \gg p$ and $N_d \gg N_a$. Thus,

$$\frac{d^2\phi(x)}{dx^2} = \frac{q}{\varepsilon_t}N_d\left\{\exp\left[\frac{q\phi(x)}{kT}\right] - 1\right\} \tag{4}$$

where $\phi(x)$ is positive, satisfying boundary conditions

$$\phi(x = \infty) = 0 \text{ and } \left.\frac{d\phi}{dx}\right|_{x=\infty} = 0$$

To simplify Eq. (4), set $u \equiv \frac{q\phi(x)}{kT}$ and introduce Debye length $L_d \equiv \sqrt{\frac{\varepsilon_t kT}{q^2 N_d}}$.

$$\frac{d^2 u}{dx^2} = \frac{1}{L_d^2}(e^u - 1)$$

Multiplying both sides by $2\frac{du}{dx}$,

$$2\frac{du}{dx}\frac{d^2 u}{dx^2} = \frac{1}{L_d^2}(e^u - 1)2\frac{du}{dx}$$

$$\frac{d}{dx}\left(\frac{du}{dx}\right)^2 = \frac{2}{L_d^2}\frac{d(e^u - u)}{dx}$$

$$\left(\frac{du}{dx}\right)^2 = \frac{2}{L_d^2}[(e^u - u) + c_1]$$

Using boundary conditions, $\left.\frac{du}{dx}\right|_{x=\infty} = \left.\frac{q}{kT}\frac{d\phi}{dx}\right|_{x=\infty} = 0$, and $u(x = \infty) = \frac{q}{kT}\phi(x = \infty) = 0$,

obtain

$$\frac{du}{dx} = -\frac{\sqrt{2}}{L_d}\sqrt{e^u - u - 1}$$

where negative sign is taken to ensure a negative slope. Thus,

$$-\frac{du}{\sqrt{\exp(u) - u - 1}} = \frac{\sqrt{2}}{L_d}dx \tag{5}$$

On the metal side, $qN_s = \varepsilon_{ox} F$. Neutrality condition in the MOT structure should be satisfied.

$$\int_0^\infty \rho(x)dx = qN_s \Rightarrow \int_0^\infty qN_d[\exp(u) - 1]dx = qN_s$$

Multiplying both sides of Eq. (5) by $qN_d[\exp(u) - 1]$ and integrating from 0 to $\infty$, obtain

$$\text{LHS} = -qN_d \int_{u_s}^0 \frac{[\exp(u) - 1]du}{\sqrt{\exp(u) - u - 1}}; \quad \text{RHS} = \frac{\sqrt{2}}{L_d}qN_s$$

$$-\int_{u_s}^{0} \frac{d[\exp(u) - u - 1]}{\sqrt{\exp(u) - u - 1}} = \frac{\sqrt{2}}{L_d} \frac{N_s}{N_d} \Rightarrow \sqrt{\exp(u_s) - u_s - 1} = \frac{1}{\sqrt{2}} \frac{N_s}{N_d L_d}$$

$$\sqrt{\exp\left(\frac{q\phi_s}{kT}\right) - \frac{q\phi_s}{kT} - 1} = \frac{1}{\sqrt{2} L_d} \frac{N_s}{N_d} \tag{6}$$

where $\phi_s = \phi(x = 0)$ is the potential at the TCO-oxide surface, and $u_s = \frac{q\phi_s}{kT}$. Solving Eq. (6), we can find the surface potential $\phi_s$. Also, we obtain the expression of the volume free charge density at the surface:

$$N_v(x = 0) = N_d \exp\left(\frac{q\phi_s}{kT}\right) = N_d \exp(u_s) \tag{7}$$

To find the potential distribution inside the TCO, integrate Eq. (5) from 0 to arbitrary $x$.

$$\frac{\sqrt{2}}{L_d} dx = -\frac{du}{\sqrt{e^u - u - 1}} \Rightarrow \frac{\sqrt{2}}{L_d} x = \int_u^{u_s} \frac{dy}{\sqrt{e^y - y - 1}}$$

$$x = \frac{L_d}{\sqrt{2}} \int_u^{u_s} \frac{dy}{\sqrt{e^y - y - 1}} \tag{8}$$

$u(x)$ and hence $\phi(x)$ can be numerically solved in Equation (5). Finally, the induced charge distribution inside the TCO can be determined through

$$N_v(x) = N_d \exp\left[\frac{q\phi(x)}{kT}\right] \tag{9}$$

4.2.1. Discussion

When $u_s \gg 1$, $\sqrt{\exp(u_s) - u_s - 1} = \frac{1}{\sqrt{2}} \frac{N_s}{N_d L_d}$. Thus, $\exp(u_s) \approx \frac{1}{2}\left(\frac{N_s}{N_d L_d}\right)^2$

$$N_v(x = 0) = N_d \exp(u_s) \approx \frac{1}{2}\left(\frac{N_s}{N_d L_d}\right)^2 N_d = \frac{1}{2}\left(\frac{N_s}{N_d}\right)^2 N_d \frac{q^2 N_d}{\varepsilon_t kT} = \frac{(qN_s)^2}{2\varepsilon_t kT}$$

$$N_v(x = 0) \approx \frac{(qN_s)^2}{2\varepsilon_t kT} \tag{10}$$

Thus, under high bias the accumulated volume carrier density at the oxide-TCO interface is proportional to square of applied voltage.

Consider the same MOT structure as discussed in previous section. If $V_b = 20$ V positive bias is applied across the HfO$_2$, the excited surface charge will be $qN_s = \varepsilon_{ox}F = 0.443$ C/m$^2$ or $N_s = 2.76 \times 10^{18}$ m$^{-2}$ according to Eq. (3). The surface potential $\phi_s = 0.161$ V, which is still considerably smaller than the bias voltage. The volume charge density at the surface will be $N_v(x=0) = 2.09 \times 10^{23}$ cm$^{-3}$, which is even higher than the electron density of gold and silver! When the surface potential $\phi_s$ is comparable to bias voltage $V_b$, the electric field across the oxide layer should be revised to $F = V_b - \phi_s$, and Eq. (6) becomes

$$\sqrt{\exp\left(\frac{q\phi_s}{kT}\right) - \frac{q\phi_s}{kT} - 1} = \frac{1}{\sqrt{2}L_d} \frac{\varepsilon_{ox}}{qN_d} \frac{V_b - \phi_s}{d} \tag{11}$$

Plug in the carrier concentration of the ITO $N_d = 4 \times 10^{20}$ cm$^{-3}$. The Debye length can be calculated as 0.215 nm. Figure 3(a) plots the charge distribution as a function of depth into ITO. As can be seen, most of charge is confined within less than 0.1 nm. In this ideal model, the charge is really surface charge. Figure 3(b) plots the complex optical dielectric constant as a function depth into the ITO layer.

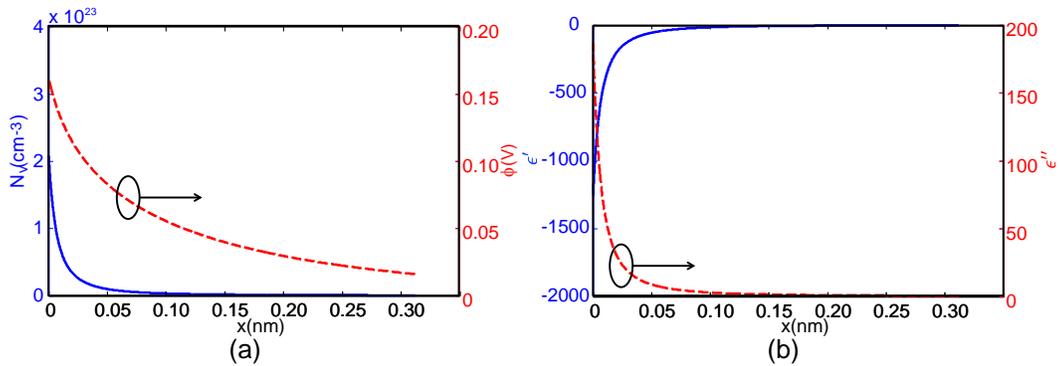

**Figure 3. (a) The volume charge and potential as a function of depth into the ITO layer. (b) The complex optical dielectric constant as a function depth into the ITO layer.**

4.2.2. Interface-Spread Effect

However, the measured accumulation layer thickness is much larger than Debye length in most cases. This was verified either by ellipsometry or attenuated total reflectance measurements [32,33]. We attribute this to be "interface-spread-effect". At the oxide-TCO interface, the material at one or both sides is amorphous. Thus, their contact is not perfect and the interface is actually a transitional range, where the TCO interface is intruded by air or oxide defects, resulting in an irregular surface instead of a plane. Ideally, the TCO surface is exactly located at $x = 0^+$ in Fig. 2(b); practically, the surface may vary from $0^+$ to $l_{ac}$. As a result, the surface charge is now spread into volume charge with thickness $l_{ac}$.

Consider a set of optical parameters for low loss ITO:

$$\epsilon_\infty = 3.9, \quad m^* = 0.35 m_e, \quad \gamma = 1.8 \times 10^{14} \text{ rad/s}$$

$$N_d = 4 \times 10^{20} \text{ cm}^{-3} \Rightarrow \omega_p = \sqrt{\frac{Nq^2}{\epsilon_o m^*}} = 1.907 \times 10^{15} \text{ rad/s}$$

The corresponding cross-over wavelength: $\lambda_{cr} = 1.986$ μm.

If $V_b = +20$ V bias is applied across the HfO$_2$ layer, and assume the accumulation layer thickness is $l_{ac} = 5$ nm, the excited charge accumulation will be

$$\Delta N = \frac{N_s}{l_{ac}} = 5.52 \times 10^{20} \text{cm}^{-3}.$$

The total charge in the accumulation layer will then be

$$N_2 = 9.52 \times 10^{20} \text{ cm}^{-3}.$$

The plasms frequency for the accumulation layer will become

$$\omega_p' = \sqrt{9.52/4}\,\omega_p = 2.942 \times 10^{15} \text{ rad/s}.$$

The corresponding cross-over wavelength: $\lambda_{cr} = 1.274$ μm. This means at wavelength longer than 1.274 μm ITO will work at its metallic state.

### 4.3. Mode 3: Epsilon-Near-Zero State

According to Eq. (1), the real-part of optical dielectric constant can be calculated as

$$\epsilon' = \epsilon_\infty - \frac{\omega_p^2}{\omega^2 + \gamma^2} \tag{12}$$

Thus, $\epsilon' = 0$ when $\omega_p = \sqrt{\epsilon_\infty(\omega^2 + \gamma^2)}$. Assume the working wavelength is exactly the cross-over wavelength, $\lambda_{cr}$=1274 nm. The corresponding imaginary part $\epsilon'' = \frac{\epsilon_\infty \gamma}{\omega} = 0.475$, which makes the magnitude of dielectric constant $|\epsilon| = 0.475$. This state is called epsilon-near-zero (ENZ) state. Optically, ENZ is a critical point, where the optical property is transiting between "dielectric state" (with positive dielectric constant) and "metallic state" (with negative dielectric constant). Any slight change in the dielectric constant may result in "optical property inversion" or the switch between these two distinct states.

It is worthy to note that the defects at the boundary may increase the scattering loss and the damping factor, $\gamma$ may vary with carrier concentration [37]. Both cases will lead to the increase of the imaginary part of the optical dielectric constant, $\epsilon''$, which may impede better realization of ENZ.

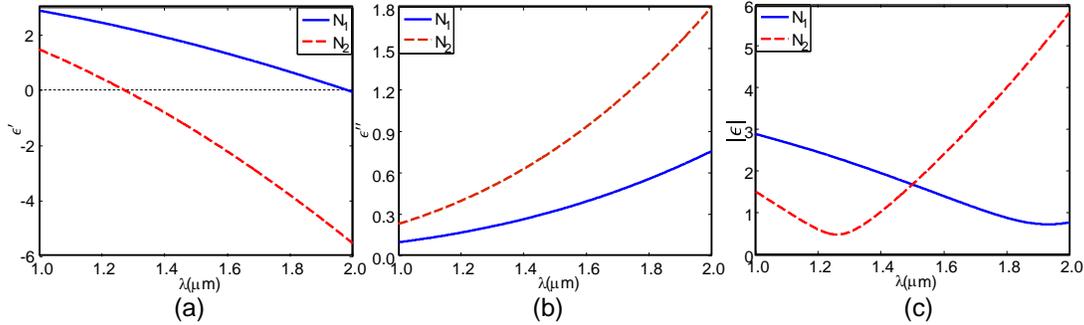

Figure 3. The comparison of ITO optical dielectric constants for no-bias and accumulation cases: (a) the real part , (b) the imaginary part, and (c) the magnitude as a function of working wavelength.

## 5. MODULATOR OPERATION PRINCIPLE

As can be seen from Fig. 3(a, b), the optical dielectric constant within the accumulation layer, both real and imaginary parts, can be greatly tuned, at $\lambda$=2000nm from -0.055+j0.756 to -5.512+j1.799, at $\lambda$=1000nm from 2.885+j0.097 to 1.483+j0.231, when the 20 V bias is applied. The light absorption in a waveguide can be determined according to $p_d = \frac{1}{2}\sigma E^2 = \frac{1}{2}\omega\epsilon''\epsilon_o E^2$. The change of $\epsilon'$ may alter the electric field distribution of $E$, whereas the change of $\epsilon''$ means the variation of $\sigma$, which is proportional to power dissipation. In addition, the magnitude of the optical dielectric constant can be also greatly tuned as shown in Fig. 3(c). Therefore, very efficient EO modulators can be achieved based the MOT structure. In this section, we will analyze the applications of optical property inversion in optical modulators.

### 5.1. Switching between Dielectric and Metallic States

The MOT devices to be considered are based on the silicon-on-insulator (SOI) technique. Fabrication can start from SOI with 120-nm p-Si on $SiO_2$. First, 10-nm $HfO_2$ is grown by atomic layer deposition (ALD), then 10-nm ITO is deposited by RF sputtering with the desired carrier concentration, $N_0 = 4 \times 10^{20}$ cm$^{-3}$, and finally 120-nm n-Si is deposited by CVD. There is no free carrier in the depletion layer and $\epsilon''$ goes to zero if the absorption and scattering loss are negligible. Figure 4(a) illustrates the construction of the waveguide. For simplicity, we only consider 2D waveguides, which is a good approximation for modulators based 3D waveguides in most case.

For C-band applications, we assume the working wavelength is $\lambda$=1550 nm. The following table is the list of optical dielectric constants for three cases, namely "no-bias", "accumulation", and "depletion". In the following discussion, including next section, the accumulation state refers to the accumulation by $V_b = -20$ V, and the depletion state refers to the depletion by $V_b = +20$ V. The "inactive thickness" refers to the layer where carrier concentration remains the same as if there is no bias.

| $\lambda$=1550 nm | No bias | Accumulation | Depletion |
|---|---|---|---|
| $\epsilon = \epsilon' + j\epsilon''$ | 1.4908 + 0.3568i | -1.8339 + 0.8493i | 3.9 |
| (Active + inactive) thickness | (0+10) nm | (5+5)nm | (6.91+3.09)nm |

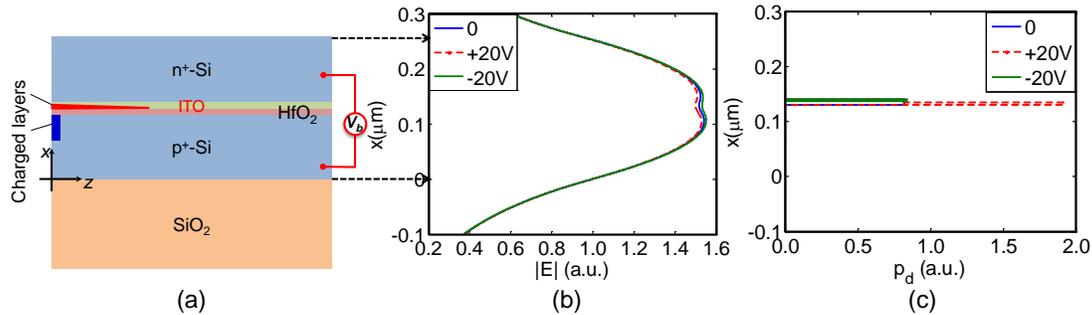

(a) (b) (c)

**Figure 4.** (a) The construction of the proposed 2D MOT structure. (b) The electric field distribution across the waveguide. (c) The power absorption across the waveguide. The thickness of the ITO film is 10nm, and oxide layer, 10 nm. The top and bottom semiconductor layers are both 120 nm thick. The refractive index of the semiconductor is assumed to be 3.47.

We solve the TE modes supported by the MOT multilayer structure based on the transfer-matrix method, and obtain the complex effective indices, $n_{\text{eff}}$, and absorption coefficient, $\alpha$, for the following three cases:

- No bias: $n_{\text{eff}} = 2.8426 + 0.0025i \Rightarrow \alpha_0 = 0.0868$ dB/μm
- Accumulation (metallic): $n_{\text{eff}} = 2.8303 + 0.0041i \Rightarrow \alpha_1 = 0.1454$ dB/μm
- Depletion (dielectric): $n_{\text{eff}} = 2.8525 + 0.0008i \Rightarrow \alpha_2 = 0.0271$ dB/μm

Thus, maximum modulation can be achieved between accumulation and depletion. A 3-dB modulator requires waveguide length $L = \frac{3}{\alpha_1} = 20.6$ μm. The corresponding insertion loss will be at least $L\alpha_2 = 0.56$ dB. Figure 4(b&c) show the electric field distribution and power absorption $p_d$ across the waveguide for the three cases.

We then consider the device operating at the TM mode, and obtain the following results:

- No bias: $n_{\text{eff}} = 1.9386 + 0.0475i \Rightarrow \alpha_0 = 1.6725$ dB/μm
- Accumulation (metallic): $n_{\text{eff}} = 2.1609 + 0.1027i \Rightarrow \alpha_1 = 3.6178$ dB/μm
- Depletion (dielectric): $n_{\text{eff}} = 2.0349 + 0.0192i \Rightarrow \alpha_2 = 0.676$ dB/μm

Similarly, maximum modulation can be achieved between accumulation and depletion. A 3-dB modulator requires waveguide length $L = \frac{3}{\alpha_1} = 0.83$ μm. The corresponding insertion loss will be at least $L\alpha_2 = 0.56$ dB.

Obviously, the MOT structure is more sensitive to TM modes. However, both TE and TM modes result in the similar level of insertion loss. If the modulator has to work for a TE mode and the device footprint is not a critical issue, the switch between dielectric and metallic states is not a bad choice.

### 5.2. Switching between Epsilon-Near-Zero and Epsilon-Far-From-Zero States

The principle was discussed in our recent work [38-41]. It is based on a new guiding mechanism ("slot waveguide") [42] and novel optical material (ENZ material), namely ENZ-slot waveguide. In a slot waveguide, electric field can be enhanced in a void slot embedded in a single mode waveguide. If the void slot is replaced by a film with refractive index even smaller than 1, resulting in an "ENZ-slot waveguide", the electric field can be further enhanced. The enhancement can only be achieved for the transverse magnetic (TM) mode, where the magnetic field is parallel to the low-index thin film, i.e. $H=H_y$ in Fig. 2(a). Due to the continuity of normal electric flux density $D_{\text{ITO},x} = D_{\text{HfO2},x}$, the dominate electric field component is inversely proportional to the slot optical dielectric constant at the HfO$_2$-ITO boundary, or

$$\epsilon_{\text{ITO}} E_{\text{ITO},x} = \epsilon_{\text{HfO2}} E_{\text{HfO2},x} \tag{13}$$

where the free charge effect is included in the complex dielectric constant. When $\epsilon_{\text{ITO}} \to 0$,

$$E_{\text{ITO},x} = \frac{\epsilon_{\text{HfO2}}}{\epsilon_{\text{ITO}}} E_{\text{HfO2},x} \to \infty. \tag{14}$$

Thus, ENZ-slot can greatly enhance the electric field in the slot, and a large portion of guided power can be confined in the slot.

On the other hand, an ENZ material always comes with absorption. Without loss of generality, we assume its optical dielectric constant to be

$$\epsilon_{\text{ITO}} = \epsilon' + j\epsilon'' = \epsilon' + j\sigma/\omega\epsilon_o \tag{15}$$

for monochromatic light ($e^{-j\omega t}$). The power absorbed by the ITO film in a unit volume,

$$p_d = \frac{1}{2}\sigma E^2 = \frac{1}{2}\omega\epsilon''\epsilon_o E^2 \propto \frac{\epsilon''}{|\epsilon_{\text{ITO}}|^2} \tag{16}$$

can be *greatly* enhanced when $\epsilon_{\text{ITO}} \to 0$, i.e. at ENZ. To maximize the absorption, the magnitude of optical dielectric constant of the slot should decrease to zero as close as possible. The absorption can be greatly modulated when $\epsilon_{\text{ITO}}$ can change between ENZ and epsilon-far-from-zero (EFFZ) states.

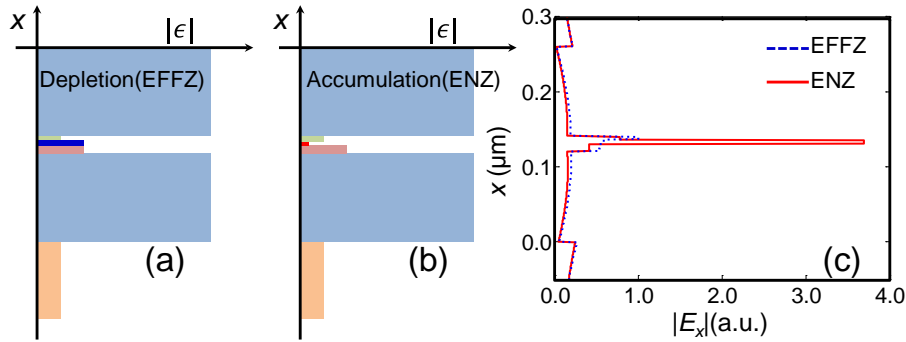

Figure 5. (a,b) The column graphs of the optical dielectric constant distribution across the waveguide when $V_b$=-20V (depletion) and $V_b$=+20V (accumulation), respectively. (c) The plots of the transverse electric field magnitude across the waveguide when $V_b$=-20V (deplation) and $V_b$=+20V (accumulation), respectively.

The modulation is more effective when the ENZ film is located at the center of the waveguide, where the magnetic field reaches its maximum, $E_z = \frac{1}{-j\omega\epsilon}\frac{\partial H_y}{\partial x} \to 0$, and electric field is mostly contributed by its normal component $E_x$.

To design a device working at ENZ, we need to choose cross-over wavelength as the working wavelength, either the no bias state is ENZ, i.e. $\lambda$=1986 nm, or accumulation state is ENZ, i.e. $\lambda$=1274 nm. The following table is the list of the parameters for working wavelength $\lambda$ =1274 nm.

| $\lambda$=1274 nm | No bias | Accumulation | Depletion |
|---|---|---|---|
| $\epsilon = \epsilon' + j\epsilon''$ | 2.2610 + 0.1995i | 0.4749i | 3.9 |
| (Active + inactive) thickness | (0+10) nm | (5+5)nm | (6.91+3.09)nm |

Figure 5(a, b) illustrates the optical dielectric constant distribution across the waveguide for the depletion and accumulation states. We solve the TM modes supported by the MOT multilayer structure and obtain the complex effective indices and absorption coefficients for the following three cases:

- No bias: $n_{\text{eff}} = 2.2819 + 0.0199i \Rightarrow \alpha_0 = 0.8503$ dB/μm
- Accumulation (ENZ): $n_{\text{eff}} = 2.1066 + 0.4880i \Rightarrow \alpha_1 = 20.9030$ dB/μm
- Depletion (dielectric): $n_{\text{eff}} = 2.3536 + 0.0071i \Rightarrow \alpha_2 = 0.3025$ dB/μm

Maximum modulation can be achieved between accumulation and depletion. A 3-dB modulator requires waveguide length $L = \frac{3}{\alpha_1} = 0.144$ μm. The corresponding insertion loss will be at least $L\alpha_2 = 0.0434$ dB. Figure 5(c) plots $|E_x|$ distribution.

The following table is the list of parameters for working wavelength $\lambda$ =1986 nm.

| $\lambda$=1986 nm | No bias | Accumulation | Depletion |
|---|---|---|---|
| $\epsilon = \epsilon' + j\epsilon''$ | -0.0015 + 0.7404i | -5.3855 + 1.7622i | 3.9 |
| (Active + inactive) thickness | (0+10) nm | (5+5)nm | (6.91+3.09)nm |

Similarly, we obtain the complex effective indices and absorption coefficients:

- No bias (ENZ): $n_{\text{eff}} = 2.0300 + 0.6170i \Rightarrow \alpha_0 = 16.9551$ dB/μm

- Accumulation (metallic): $n_{\text{eff}} = 2.3634 + 0.5290i \Rightarrow \alpha_1 = 14.5368$ dB/μm
- Depletion (dielectric): $n_{\text{eff}} = 2.3634 + 0.2690i \Rightarrow \alpha_2 = 7.3921$ dB/μm

Maximum modulation can be achieved between no-bias and depletion. A 3-dB modulator requires waveguide length $L = \frac{3}{\alpha_0} = 0.177$ μm. The corresponding insertion loss will be at least $L\alpha_2 = 1.31$ dB.

Obviously, the first case results in higher modulation depth and lower insertion loss. To operate the device in this mode, it is better to choose the ENZ state as the final state.

## 6. Summary

The metal-oxide-TCO, i.e. MOT, structure provides an effective approach to realize the photonic counterparts of electronic MOSFETs. When bias is applied across the MOT, the charged layer in the TCO may result in "optical property inversion", where the optical property can be switched between dielectric and metallic states, or between EFFZ and ENZ states. The theory and example for the MOT structure for EO modulation were discussed. MOT based on "optical property inversion" may provide new vision for the development of nanophotonic devices.

The publication was made possible by Grant Number W911NF-12-1-0451 from the United States Army. This material is based upon work partially supported by the National Science Foundation under Award No. ECCS-1308197.